\documentclass[12pt]{article}

\makeatletter
\if@twoside \oddsidemargin -17pt \evensidemargin 00pt
\else \oddsidemargin 00pt \evensidemargin 00pt
\fi
\expandafter\ifx\csname mathrm\endcsname\relax\def\mathrm#1{{\rm #1}}\fi

\newcounter{saveeqn}

\@addtoreset{equation}{section}

\makeatother

\def\beq{\begin{equation}}
\def\eeq{\end{equation}}
\def\beqar{\begin{eqnarray}}
\def\eeqar{\end{eqnarray}}
\def\barr#1{\begin{array}{#1}}
\def\earr{\end{array}}
\def\bfi{\begin{figure}}
\def\efi{\end{figure}}
\def\btab{\begin{table}}
\def\etab{\end{table}}
\def\bce{\begin{center}}
\def\ece{\end{center}}

\def\text{\textstyle}

\def\de{\delta}

\def\si{\sigma}

\def\refeq#1{\mbox{(\ref{#1})}}
\def\reffi#1{\mbox{Fig.~\ref{#1}}}

\def\citere#1{\mbox{Ref.~\cite{#1}}}
\def\citeres#1{\mbox{Refs.~\cite{#1}}}

\def\solid{\raise.9mm\hbox{\protect\rule{1.1cm}{.2mm}}}
\def\dash{\raise.9mm\hbox{\protect\rule{2mm}{.2mm}}\hspace*{1mm}}

\newcommand{\GeV}{\unskip\,\mathrm{GeV}}
\newcommand{\MeV}{\unskip\,\mathrm{MeV}}

\def\mathswitchr#1{\relax\ifmmode{\mathrm{#1}}\else$\mathrm{#1}$\fi}

\newcommand{\PV}{\mathswitch V}
\newcommand{\PW}{\mathswitchr W}
\newcommand{\PZ}{\mathswitchr Z}

\newcommand{\Pe}{\mathswitchr e}

\newcommand{\Pep}{\mathswitchr {e^+}}
\newcommand{\Pem}{\mathswitchr {e^-}}

\newcommand{\PWp}{\mathswitchr {W^+}}
\newcommand{\PWm}{\mathswitchr {W^-}}
\newcommand{\PWpm}{\mathswitchr {W^\pm}}

\def\mathswitch#1{\relax\ifmmode#1\else$#1$\fi}

\newcommand{\rd}{{\mathrm{d}}}

\newcommand{\born}{{\mathrm{Born}}}
\newcommand{\nf}{{\mathrm{nf}}}

\def\Re{\mathop{\mathrm{Re}}\nolimits}

\def\draftdate{\relax}
\def\mda{\relax}
\def\mua{\relax}
\def\mla{\relax}
\def\draft{
\def\thtystars{******************************}
\def\sixtystars{\thtystars\thtystars}
\typeout{}
\typeout{\sixtystars**}
\typeout{* Draft mode!
         For final version remove \protect\draft\space in source file *}
\typeout{\sixtystars**}
\typeout{}
\def\draftdate{\today}
\def\mua{\marginpar[\boldmath\hfil$\uparrow$]%
                   {\boldmath$\uparrow$\hfil}%
                    \typeout{marginpar: $\uparrow$}\ignorespaces}
\def\mda{\marginpar[\boldmath\hfil$\downarrow$]%
                   {\boldmath$\downarrow$\hfil}%
                    \typeout{marginpar: $\downarrow$}\ignorespaces}
\def\mla{\marginpar[\boldmath\hfil$\rightarrow$]%
                   {\boldmath$\leftarrow $\hfil}%
                    \typeout{marginpar: $\leftrightarrow$}\ignorespaces}
\def\Mua{\marginpar[\boldmath\hfil$\Uparrow$]%
                   {\boldmath$\Uparrow$\hfil}%
                    \typeout{marginpar: $\Uparrow$}\ignorespaces}
\def\Mda{\marginpar[\boldmath\hfil$\Downarrow$]%
                   {\boldmath$\Downarrow$\hfil}%
                    \typeout{marginpar: $\Downarrow$}\ignorespaces}
\def\Mla{\marginpar[\boldmath\hfil$\Rightarrow$]%
                   {\boldmath$\Leftarrow $\hfil}%
                    \typeout{marginpar: $\Leftrightarrow$}\ignorespaces}
\overfullrule 5pt
\oddsidemargin -15mm
\marginparwidth 29mm
}

\makeatletter

\def\eqnarray{\stepcounter{equation}\let\@currentlabel=\theequation
\global\@eqnswtrue
\global\@eqcnt\z@\tabskip\@centering\let\\=\@eqncr
$$\halign to \displaywidth\bgroup\hskip\@centering
  $\displaystyle\tabskip\z@{##}$\@eqnsel&\global\@eqcnt\@ne
  \hskip 2\arraycolsep \hfil${##}$\hfil
  &\global\@eqcnt\tw@ \hskip 2\arraycolsep $\displaystyle\tabskip\z@{##}$\hfil
   \tabskip\@centering&\llap{##}\tabskip\z@\cr}
\def\appendix{\par
 \setcounter{section}{0} \setcounter{subsection}{0}
 \def\thesection{\Alph{section}}}

\makeatother

\begin{document}

\thispagestyle{empty}
\def\thefootnote{\fnsymbol{footnote}}
\setcounter{footnote}{1}
\null
\draftdate\hfill CERN-TH/98-72 \\
\strut\hfill PSI-PR-98-07\\
\strut\hfill hep-ph/9803306
\vskip 0cm
\vfill
\begin{center}
{\Large \bf
Further numerical results \\
on non-factorizable corrections to \boldmath{$\Pep\Pem\to 4\,$fermions}
\par} \vskip 2.5em
{\large
{\sc A.~Denner%
}\\[1ex]
{\normalsize \it Paul Scherrer Institut, W\"urenlingen und Villigen\\
CH-5232 Villigen PSI, Switzerland}\\[2ex]
{\sc S.~Dittmaier%
}\\[1ex]
{\normalsize \it Theory Division, CERN\\
CH-1211 Geneva 23, Switzerland}\\[2ex]
{\sc M. Roth%
}\\[1ex]
{\normalsize \it Paul Scherrer Institut, W\"urenlingen und Villigen\\
CH-5232 Villigen PSI, Switzerland\\
and\\
Institut f\"ur Theoretische Physik, ETH-H\"onggerberg\\
CH-8093 Z\"urich, Switzerland
}\\[2ex]
}
\par \vskip 1em
\end{center}
\par
\vskip .0cm 
\vfill 
{\bf Abstract:} \par 
Numerical results on non-factorizable corrections to
$\Pep\Pem\to\PW\PW\to 4\,$fermions with semi-leptonic and hadronic
final states, as well as to $\Pep\Pem\to\PZ\PZ\to 4\,$fermions, 
are presented.  The corrections turn out to be small in comparison to the
experimental uncertainty of LEP2, but they might
compete with the expected accuracy at
future $\Pep\Pem$-colliders.
\par
\vskip 1cm
\noindent
March 1998 
\par
\null
\setcounter{page}{0}
\clearpage
\def\thefootnote{\arabic{footnote}}
\setcounter{footnote}{0}

In order to match the experimental accuracy of roughly 1\% at LEP2,
the precision of the predictions for the cross sections of
$\Pep\Pem\to 4\,$fermions should be at, or rather exceed the per-cent
level.  This precision requires us to go beyond the narrow-width
approximation for \PW- and \PZ-boson-pair production as well as to
include electroweak radiative corrections.  For general aspects and
further details of the strategy for such precision calculations we
refer to review articles \cite{wwrev}.

The full electroweak ${\cal O}(\alpha)$ corrections to 
$\Pep\Pem\to\PW\PW,\PZ\PZ\to 4\,$fermions are not known
and will not be available in the near future.
A few decay widths above the threshold for 
gauge-boson-pair production 
these corrections should be sufficiently well described by
the so-called double-pole approximation, which consists of taking 
into account only those contributions that are enhanced by two resonant 
W or Z~bosons. In this approximation the ${\cal O}(\alpha)$ corrections 
can be separated into {\it factorizable} and {\it non-factorizable} 
corrections. The former can be associated either with the
gauge-boson-pair
production or the gauge-boson decay subprocesses and are well-known.
The latter include the corrections in which the two W- or Z-boson 
resonances
are not independent; they are due to the exchange of soft photons
between the different subprocesses. The non-factorizable corrections
have recently been discussed in the literature \cite{Me96,Be97,De97}.

In this letter we supplement our results on non-factorizable
corrections presented in \citere{De97}, which agree%
\footnote{For more details of the comparison with the results of
  \citeres{Me96,Be97} we refer to \citere{De97}.}  
analytically as well as numerically with those of \citere{Be97}.  We
start by briefly recalling the salient features of the
non-factorizable corrections.  Already before their explicit
calculation it was known \cite{Fa94} that they are non-vanishing only
if not both invariant masses of the W or Z~bosons are integrated over,
i.e., for instance, that they do not influence total cross sections.

In \citere{De97} we explained why the actual form of the
non-factorizable corrections is non-universal in the sense that it
depends on the choice for the parametrization of phase space for the
real photonic corrections.  We have adopted the usual choice and taken
the invariant masses of decay fermion pairs as independent variables.
Since all corrections that are related to the initial state drop out,
the corrections neither depend on the gauge-boson-pair
production angle nor on the initial state itself. 

The non-factorizable corrections to the processes
\beq\label{process}
\Pep(p_+) + \Pem(p_-) \;\to\; \PV_1(k_1+k_2) + \PV_2(k_3+k_4)
\;\to\; f_1(k_1) + \bar f_2(k_2) + f_3(k_3) + \bar f_4(k_4)
\eeq
($\PV_1\PV_2=\PW\PW,\PZ\PZ$) 
are proportional to the lowest-order differential cross section: 
\beq
\rd \si_\nf = \de_\nf \, \rd\si_\born,
\eeq
and the relative correction factor can be written as \cite{De97}
\beq\label{nfcorrfac}
\delta_\nf(k_1,k_2;k_3,k_4) = 
\sum_{a=1,2} \, \sum_{b=3,4} \, (-1)^{a+b+1} \, Q_a Q_b \,
\frac{\alpha}{\pi} \, \Re\{\Delta(k_1+k_2,k_a;k_3+k_4,k_b)\},
\eeq
where $Q_i$ $(i=1,2,3,4)$ denotes the relative charge of fermion $f_i$. 
The function $\Delta$ is explicitly given in
\citere{De97} and has the symmetry
\beq\label{symmetry}
\Delta(k_1+k_2,k_a;k_3+k_4,k_b)=\Delta(k_3+k_4,k_b;k_1+k_2,k_a).
\eeq

In all numerical evaluations up to now,
only the purely leptonic process
$\Pep\Pem\to\PW\PW\to\nu_\ell \ell^+\ell^{\prime-}\nu_{\ell'}$ has been
considered. In this case the non-factorizable corrections are
of the order of 1\% in the LEP2 energy range, but rapidly tend to zero
for higher energies. The corrections to the single-invariant-mass
distributions $\rd\sigma/\rd M_\pm$ are identical and shift the peaks of
the distributions by an amount of $1$--2$\MeV$ 
for typical LEP2 energies, which is in fact negligible at LEP.

The invariant-mass distributions for the hadronic decay channels are
of particular importance for the reconstruction of the W-boson mass
from the W-boson decay products.
The non-factorizable corrections to the invariant-mass distributions 
are different for different final states and in general also for the
intermediate \PWp~ and \PWm~bosons.%
\footnote{%
In \citere{Be97} and in the preprint version of \citere{De97}
it has been argued that
the relative non-factorizable corrections to pure invariant-mass
distributions are identical for all final states in
$\Pep\Pem\to\PW\PW\to 4\,$fermions and vanish for Z-pair-mediated
four-fermion production. This was deduced from the assumption that
(up to charge factors)
the non-factorizable corrections become symmetric under the separate
interchanges $k_1\leftrightarrow k_2$ and $k_3\leftrightarrow k_4$
after integration over all decay angles. 
Although the function $\Delta$ for the relative correction has this
property,
this assumption is
not correct, because the differential lowest-order cross section is not
symmetric under these interchanges.}
The invariant-mass distributions to the intermediate $\PWpm$~bosons 
coincide only if the complete process is CP-symmetric. In this context,
CP symmetry does not distinguish between the different fermion generations,
since we work in double-pole approximation and neglect fermion masses;
in other words, the argument also applies to final states like 
$\nu_\Pe\Pep\mu^-\bar\nu_\mu$ and $u\bar d s\bar c$, which are not
CP-symmetric in the strict sense.
Thus, we end up with equal distributions for the $\PWpm$~bosons in the purely
leptonic and purely hadronic channels, respectively, but not in the 
semi-leptonic case. 

\bfi
\centerline{
\setlength{\unitlength}{1cm}
\begin{picture}(16,7.8)
\put(0,0){\includegraphics{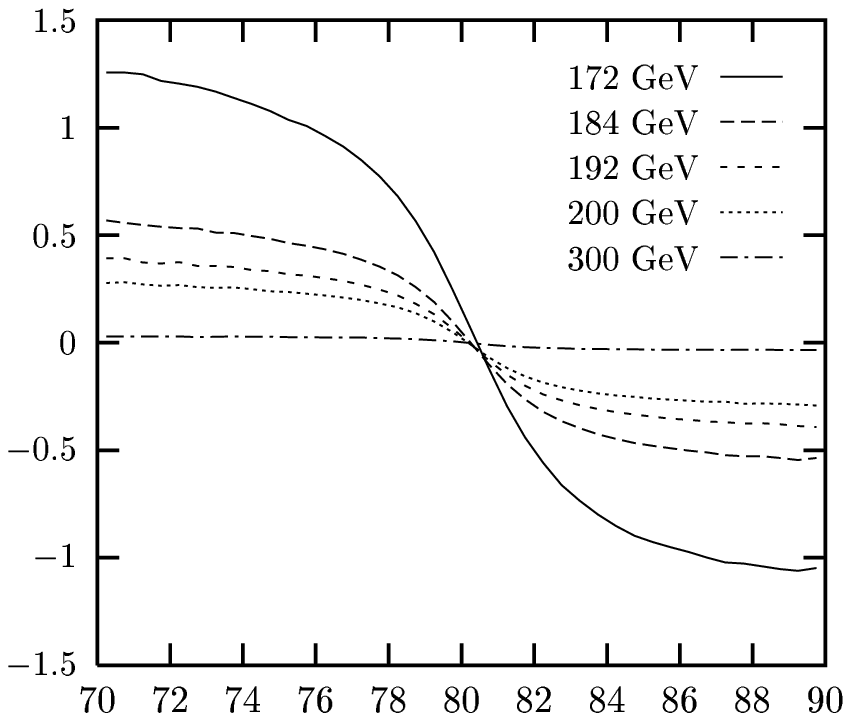}}
\put(0,0){\includegraphics{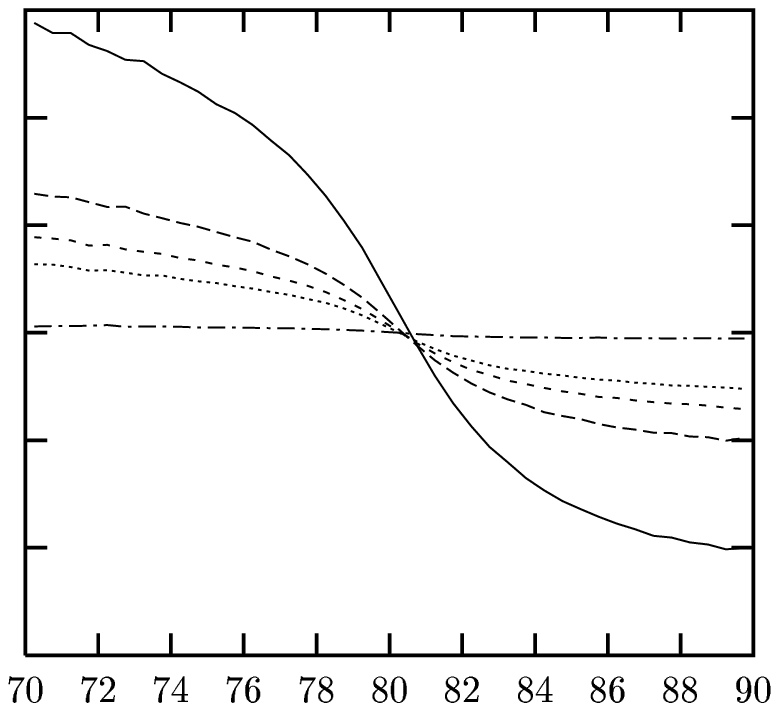}}
\put(0.0,5.2){\makebox(1,1)[c]{$\de_\nf/\%$}}
\put(4.7,-0.3){\makebox(1,1)[cc]{{$M_\pm/{\rm GeV}$}}}
\put(11.7,-0.3){\makebox(1,1)[cc]{{$M_\pm/{\rm GeV}$}}}
\put( 2.8,7.3){$\Pep\Pem\to\PW\PW\to\nu_\ell \ell^+\ell^{\prime-}\nu_{\ell'}$}
\put(10.0,7.3){$\Pep\Pem\to\PW\PW\to u\bar d d'\bar u'$}
\end{picture}
}
\vspace*{3mm}
\centerline{
\setlength{\unitlength}{1cm}
\begin{picture}(16,7.8)
\put(0,0){\includegraphics{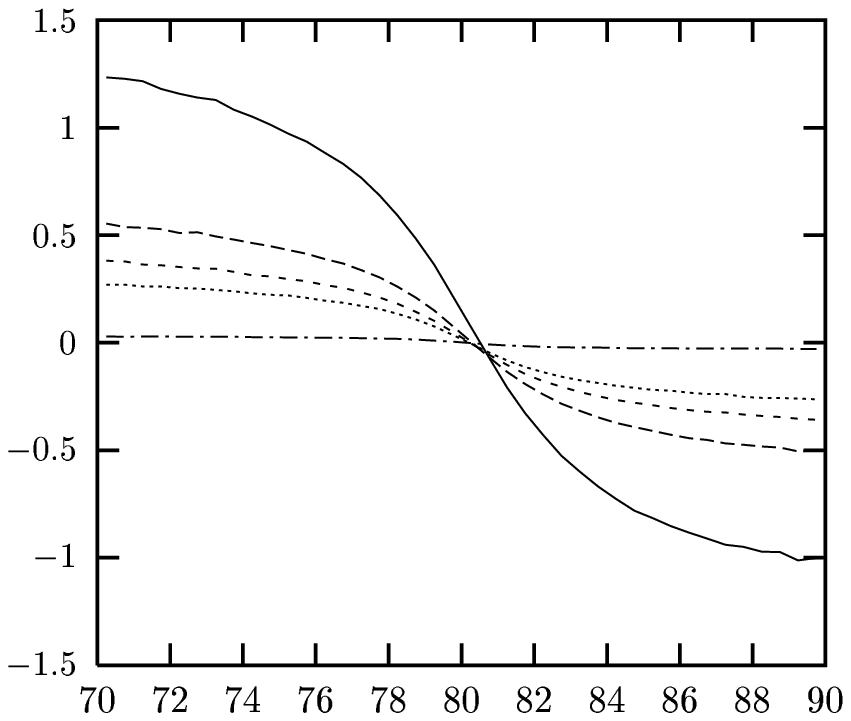}}
\put(0,0){\includegraphics{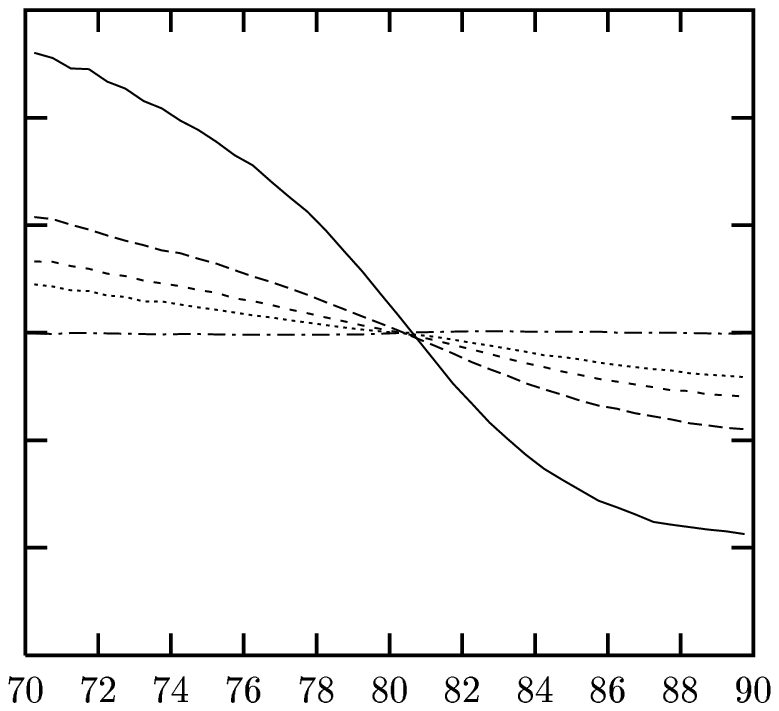}}
\put(0.0,5.2){\makebox(1,1)[c]{$\de_\nf/\%$}}
\put(4.7,-0.3){\makebox(1,1)[cc]{{$M_+/{\rm GeV}$}}}
\put(11.7,-0.3){\makebox(1,1)[cc]{{$M_-/{\rm GeV}$}}}
\put( 2.8,7.3){$\Pep\Pem\to\PW\PW\to\nu_\ell \ell^+ d\bar u$}
\put(10.0,7.3){$\Pep\Pem\to\PW\PW\to\nu_\ell \ell^+ d\bar u$}
\end{picture}
}
\caption[]{Relative non-factorizable corrections to the single-%
invariant-mass distributions $\protect\rd\sigma/\protect\rd M_\pm$ 
for $\Pep\Pem\to\PW\PW\to 4\,$fermions with different final states
for various centre-of-mass energies.}
\label{Winvmass}
\efi
Figure~\ref{Winvmass} shows the non-factorizable corrections to 
the single-invariant-mass distributions for leptonic,
hadronic, and semi-leptonic final states at various centre-of-mass
energies%
\footnote{For details concerning the implementation of the correction
factor $\de_\nf$ in {\tt EXCALIBUR} \cite{Be94} and for the input
parameters we refer to \citere{De97}.
}%
. We observe the same qualitative features for all final states;
the corrections are positive below resonance and negative above. 
Quantitatively the differences between the corrections to the different
final states are small; we note that the slopes of the
corrections on resonance, which are responsible for the shift in the
maximum of the distribution, are maximal for the leptonic final state.
Therefore, we conclude that the W-boson mass determination by
invariant-mass reconstruction at LEP2 is not significantly influenced by
non-factorizable corrections.

The authors of \citere{Be97} have also calculated \cite{Be98} the 
non-factorizable
corrections to the single-invariant-mass distributions shown in
\reffi{Winvmass} for $\sqrt{s}=172\GeV$ and $184\GeV$. They find good
agreement with our results for positive invariant masses. However,
their corrections are antisymmetric and therefore differ from our
results for negative invariant masses. The differences are of the
order of non-doubly-resonant corrections and due to
different parametrizations of the corrections. For a discussion of
these differences we refer to \citere{De97}.

In our discussion of the non-factorizable corrections to
$\Pep\Pem\to\PW\PW\to 4\,$leptons in \citere{De97} we also investigated 
their influence on various angular and energy distributions with fixed
invariant masses for the final-state fermion pairs. We have repeated
this analysis for hadronic and semi-leptonic final states and found
corrections of the same order of magnitude, viz.\ of typically 1\% 
at LEP2 energies.

For the production channels via a resonant Z-boson pair,
$\Pep\Pem\to\PZ\PZ\to 4\,$fermions, we have $f_1=f_2$ and $f_3=f_4$.
Owing to Bose symmetry the lowest-order cross section $\rd\si_\born$ 
is invariant under the set of interchanges
$(k_1,k_2)\leftrightarrow(k_3,k_4)$. 
This symmetry, which is respected by the non-factorizable corrections
[see \refeq{symmetry}],
implies that the single-invariant-mass distributions to each of the 
final-state fermion pairs of the two Z-boson decays are equal.
CP invariance leads to the additional symmetry with respect to
$(p_+,k_1,k_2)\leftrightarrow(p_-,k_4,k_3)$; after integration over the
Z-pair production angle this substitution reduces to
$(k_1,k_2)\leftrightarrow(k_4,k_3)$.
In view of non-factorizable corrections it is also interesting to
inspect the behaviour of $\rd\si_\born$ under the replacements 
$k_1\leftrightarrow k_2$ and $k_3\leftrightarrow k_4$ separately, since
terms in $\rd\si_\born$ that are symmetric in at least one of these
substitutions do not contribute to $\rd\si_\nf$ 
if all decay angles are integrated over.
This is a direct consequence of the antisymmetry of $\de_\nf$ in  
each of the substitutions 
$k_1\leftrightarrow k_2$ and $k_3\leftrightarrow k_4$, which follows
from \refeq{nfcorrfac} and $Q_1=Q_2$, $Q_3=Q_4$.
 
In order to study the behaviour of $\rd\si_\born$ under the
replacements $k_1\leftrightarrow k_2$ and $k_3\leftrightarrow k_4$, it
is convenient to consider the helicity amplitudes for the two signal
diagrams for $\Pep\Pem\to\PZ\PZ\to 4\,$fermions, which contain two
resonant Z-boson propagators. These amplitudes are proportional to the
right- and left-handed couplings $g_i^\pm=v_i\mp a_i$ of each fermion
$f_i=f_1,f_3$ to the Z~boson.  As can be seen from the explicit form
of the amplitudes, the substitution $k_1\leftrightarrow k_2$
transforms the helicity amplitudes to those with reversed helicities
of the fermions $f_1$ and $\bar f_2=\bar f_1$ apart from changing the
couplings $g_1^\pm$ into $g_1^\mp$.  Therefore, the differential
lowest-order cross section, i.e.\ the squared helicity amplitudes
summed over all final-state polarizations, can be split into two
parts: one is symmetric in $k_1\leftrightarrow k_2$ and proportional
to $[(g_1^+)^2+(g_1^-)^2]/2= v_1^2+a_1^2$, the other is anti-symmetric
and proportional to $[(g_1^-)^2-(g_1^+)^2]/2=2v_1 a_1$.  The analogous
reasoning applies to the substitution $k_3\leftrightarrow k_4$. After
performing the angular integrations, we finally find that the
lowest-order cross section is proportional to
$(v_1^2+a_1^2)(v_3^2+a_3^2)$, and the non-factorizable correction
proportional to $4Q_1 v_1 a_1 Q_3 v_3 a_3$, where the charge factors
$Q_i$ stem from the correction factor $\de_\nf$. Comparing pure
invariant-mass distributions for different final states, the ratios of
the non-factorizable corrections should be of the same order of
magnitude as the ratios
of the corresponding coupling factors,
\beq
F = \left|\frac{4Q_1 v_1 a_1 Q_3 v_3 a_3}{(v_1^2+a_1^2)(v_3^2+a_3^2)}\right|.
\label{eq:F}
\eeq
The factors $F$ take the following values:
\beq
\arraycolsep 8pt
\begin{array}{|c||c|c|c|c|c|c|}
\hline
f_1 f_3 & \ell\ell & \ell u & \ell d & uu   & ud   & dd 
\\ \hline
F &       0.04     & 0.09   & 0.06   & 0.21 & 0.14 & 0.10
\\ \hline
\end{array}
\label{eq:Ftab}
\eeq
where $\ell$, $u$, $d$ generically refer to leptons, up-type quarks and
down-type quarks, respectively. The reason for the
smallness of the factors $F$ is different for leptons and quarks: for
leptons the suppression is due to the small coupling $v_i$ 
to the vector current, for quarks the factor $F$ is reduced by the 
relative charges $Q_i$. 

\bfi
\centerline{
\setlength{\unitlength}{1cm}
\begin{picture}(9,7.5)
\put(0,0){\includegraphics{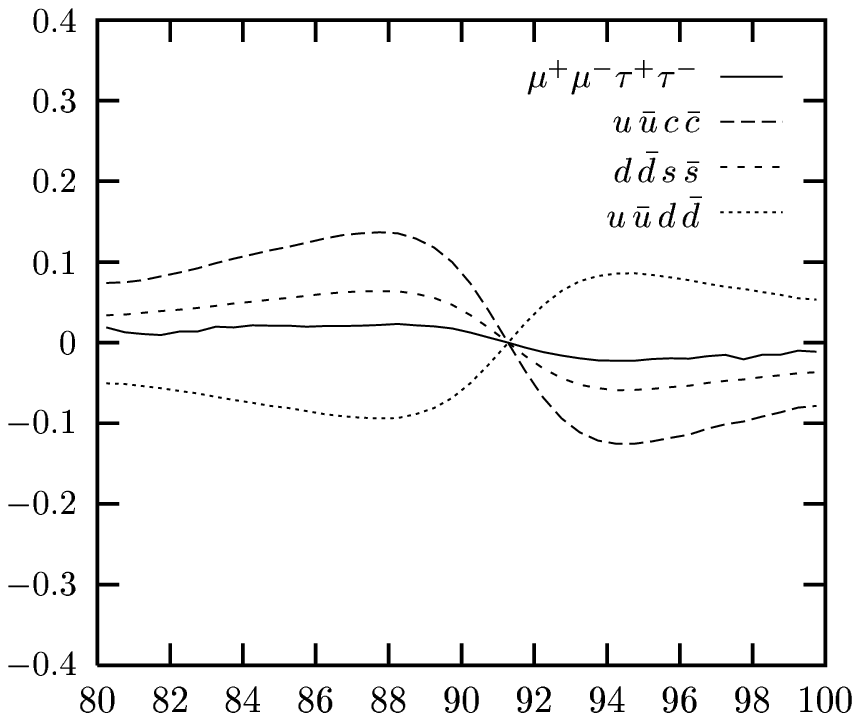}}
\put(0.0,5.2){\makebox(1,1)[c]{$\de_\nf/\%$}}
\put(4.7,-0.3){\makebox(1,1)[cc]{{$M_{1,2}/{\rm GeV}$}}}
\end{picture}
}
\caption[]{Relative non-factorizable corrections to the single-%
invariant-mass distributions $\protect\rd\sigma/\protect\rd M_{1,2}$ 
for $\Pep\Pem\to\PZ\PZ\to 4\,$fermions with different final states
for $\sqrt{s}=192\GeV$.}
\label{Zinvmass}
\efi
Figure~\ref{Zinvmass} shows the non-factorizable corrections to the
single-invariant-mass distributions $\rd\sigma/\rd M_{1,2}$, where
$M_{1,2}$ denote the invariant masses of the first and second
fermion--anti-fermion pairs, respectively. The ratios of the different curves 
are indeed of the order of magnitude of the ratios of the factors $F$ given in
\refeq{eq:Ftab}. 
For equal signs of $Q_1$ and $Q_3$ the shape of the corrections is
similar to the shape of the corrections to $\Pep\Pem\to\PW\PW\to
4\,$fermions, for opposite signs of $Q_1$ and $Q_3$ the 
shape is reversed.
The corrections by themselves are very small and
phenomenologically unimportant. The smallness of these corrections can
be qualitatively understood by comparing the factors $F$ of
$\refeq{eq:F}$ for the ratios of the couplings with the corresponding one
for the W-pair-mediated processes. For $\Pep\Pem\to\PW\PW\to 4\,$leptons
we simply have $F=1$, because in the LEP2 energy range the purely
left-handed $t$-channel diagram dominates the cross section, and no
systematic compensations are induced by symmetries. Therefore, the
factors in $\refeq{eq:Ftab}$ should directly give an estimate for the
suppression of $\de_\nf$ for $\Pep\Pem\to\PZ\PZ\to 4\,$fermions with
respect to four-lepton production via a W-boson pair. Comparing the
corrections for energies with the same distance from the respective
on-shell pair-production thresholds, i.e.\ the curve for
$\sqrt{s}=184\GeV$ in the W-boson case (\reffi{Winvmass}) with the
curves for $\sqrt{s}=192\GeV$ in the Z-boson case (\reffi{Zinvmass}), we
find reasonable agreement with our expectation. 
The authors of \citere{Be97} have reproduced the corrections shown in
\reffi{Zinvmass} with good agreement \cite{Be98}.

\bfi
\centerline{
\setlength{\unitlength}{1cm}
\begin{picture}(16,7.5)
\put(0,0){\includegraphics{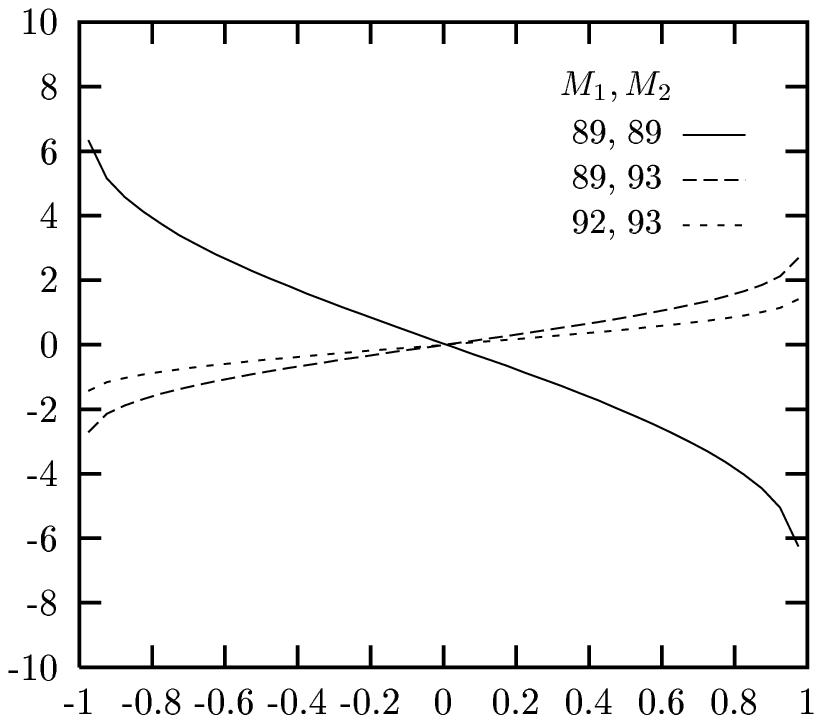}}
\put(0,0){\includegraphics{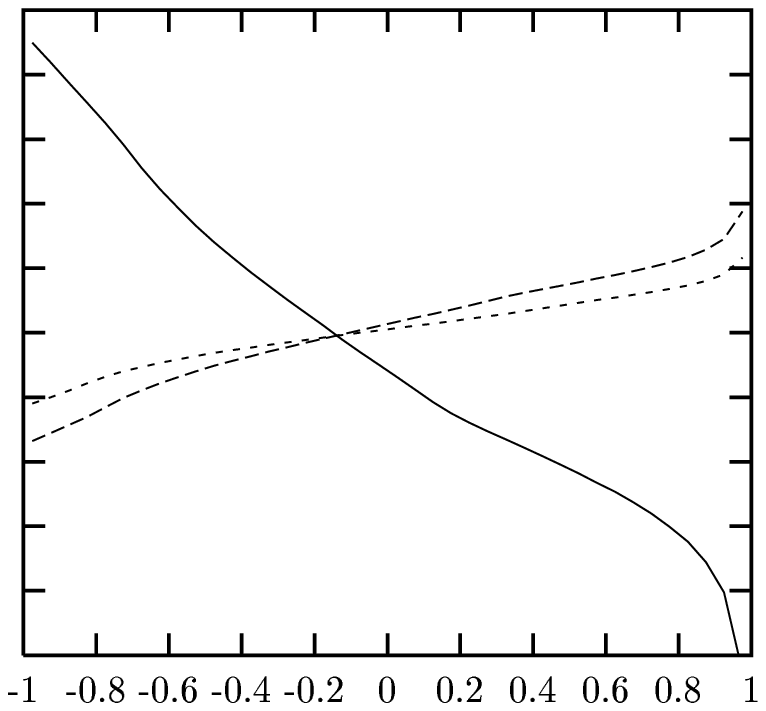}}
\put(0.0,5.2){\makebox(1,1)[c]{$\de_\nf/\%$}}
\put(4.7,-0.3){\makebox(1,1)[cc]{{$\cos\phi$}}}
\put(11.7,-0.3){\makebox(1,1)[cc]{{$\cos\theta_{\mu^+\tau^-}$}}}
\end{picture}
}
\caption[]{Relative non-factorizable corrections to the angular distributions 
$\protect\rd\sigma/\protect\rd M_1\protect\rd M_2\protect\rd\cos\phi$ and
$\protect\rd\sigma/\protect\rd M_1\protect\rd M_2\protect
\rd\cos\theta_{\mu^+\tau^-}$ 
in $\Pep\Pem\to\PZ\PZ\to\mu^-\mu^+\tau^-\tau^+$ 
for fixed values of the invariant masses $M_{1,2}$ and $\sqrt{s}=192\GeV$.}
\label{Zangles}
\efi
Finally, we inspect the impact of non-factorizable corrections to some
angular distributions in Z-pair-mediated four-fermion production for fixed 
values of the invariant masses $M_{1,2}$.
Since the presence of the suppression factor $F$ relies
on the assumption that the
phase-space integration is symmetric under 
$k_1\leftrightarrow k_2$ and $k_3\leftrightarrow k_4$, this suppression
in general does not apply to angular distributions. However, 
partial suppressions occur, e.g., if the integration is still symmetric 
under one of these substitutions and, in particular, for quarks in the 
final state because of their smaller charges.
Two examples for angular distributions without any suppression 
are illustrated in \reffi{Zangles} for the purely leptonic final state 
$\mu^-\mu^+\tau^-\tau^+$. The angle $\phi$ is defined by the two planes 
spanned by the momenta of the two fermion pairs in which the Z~bosons decay,
\newcommand{\bk}{{\bf k}}
\beq
\cos\phi = \frac{(\bk_1\times \bk_2)(\bk_3\times \bk_4)}
{|\bk_1\times \bk_2||\bk_3\times \bk_4|},
\label{eq:phi}
\eeq
and $\theta_{\mu^+\tau^-}$ denotes the angle between the momenta of the
$\mu^+$ and the $\tau^-$, respectively. The shapes of the curves in
\reffi{Zangles}, specifically the curves for the distribution in
$\cos\phi$, nicely reflect the approximate anti-symmetric behaviour in the
angular dependence, which leads to the suppression in the invariant-mass
distributions. The size of the corrections turns out to be at the level 
of a few per cent, i.e.\ they are not necessarily negligible in precision 
predictions. Note, however, that the cross section for \PZ-pair
production is only one tenth of the \PW-pair production cross section.

In conclusion, we find that the non-factorizable corrections to \PW-pair
production are not identical but similar for all final states and 
negligible compared to the LEP2 accuracy. 
In the case of \PZ-pair production, the corrections
are smaller for the invariant-mass distribution, but larger for some
angular distributions. 
In view of the smallness of the cross section for \PZ-pair production,
even these relatively large corrections are not relevant for LEP2.
The size of the non-factorizable corrections might, however, compete
with the expected experimental accuracy of future $\Pep\Pem$-colliders
with higher luminosity.

\section*{Acknowledgements}

We thank W.~Beenakker, F.A.~Berends, and A.P.~Chapovsky for discussions 
and for performing a comparison with our numerical results 
on the invariant-mass distributions.

\end{document}